\documentclass{aa}
\input epsf
\input psfig
\epsfverbosetrue

\begin{document}
%\thesaurus{09.03.2, 09.09.1, 13.07.2}

\title{Detection of $\gamma$-ray lines from interstellar $^{60}$Fe by the
high resolution spectrometer SPI}

\institute{
Centre d'\'{E}tude Spatiale des Rayonnements, B.P. N$^{o}$ 4346, 31028 Toulouse
Cedex 4, France
\and
IASF, Via E. Bassini 15, 20133 Milano, Italy
\and 
Max-Planck-Institut f\"{u}r extraterrestrische Physik, Postfach 1603, 85740
Garching, Germany
\and
DSM/DAPNIA/SAp, CEA Saclay, 91191 Gif-sur-Yvette, France
	}

\author{M.J.~Harris\inst{1} \and J.~Kn\"{o}dlseder\inst{1}
\and P.~Jean\inst{1} \and E.~Cisana\inst{2} \and R.~Diehl\inst{3}
\and G.G.~Lichti\inst{3} \and J.-P.~Roques\inst{1} \and S. Schanne\inst{4} 
\and G.~Weidenspointner\inst{1} }

\offprints{M.J. Harris}
\date{Received }

\authorrunning{M.J. Harris et al.}

\titlerunning{SPI detection of $^{60}$Fe}

\abstract{ It is believed that core-collapse supernovae (CCSN), occurring at 
a rate $\sim$once per century, have seeded the interstellar medium with
long-lived radioisotopes such as $^{60}$Fe (half-life 1.5 Myr),
which can be detected by the $\gamma$-rays emitted when they
$\beta$-decay.  Here we report the detection of the $^{60}$Fe decay 
lines at 1173 keV and 1333 keV with fluxes $3.7 \pm 1.1 \times 10^{-5} 
~\gamma ~\hbox{cm}^{-2} \hbox{s}^{-1}$ per line, in spectra taken 
by the SPI spectrometer on board
{\em INTEGRAL\/} during its first year.  The same analysis applied to the 1809 keV line of
$^{26}$Al yielded a line flux ratio $^{60}$Fe/$^{26}$Al = $0.11 \pm 0.03$.
This supports the hypothesis that there is an extra source of $^{26}$Al 
in addition to CCSN.
}

\maketitle

\section{Introduction}

The radioactive isotopes $^{26}$Al and $^{60}$Fe are both believed to be produced
in massive stars that end their lives as core collapse supernovae (CCSN; masses $>8 M_{\sun}$).
Further, they have similar half-lives ($7.4 \times
10^{5}$ yr and $1.5 \times 10^{6}$ yr respectively) which are much longer than the characteristic
interval between CCSN ($\sim 100$ yr).  Therefore they will
accumulate in the interstellar medium until a steady state is reached, and
indeed this steady-state abundance of $^{26}$Al has been detected via a flux
$\sim 4 \times 10^{-4} ~\gamma ~ \hbox{cm}^{-2} \hbox{s}^{-1}$ 
in the 1809 keV line from its $\beta$-decay (Mahoney et al. 1982, Diehl 2001).

The sky distribution of $^{26}$Al has been mapped (Kn\"{o}dlseder et al. 1999)
and, as expected, the line emission
is dominated by the Galactic plane, where massive stars are found.
It might be expected that the distribution of $^{60}$Fe line emission
(at 58 keV, 1173 keV and 1333 keV) would be very similar.  However (as we will see
in \S 4) there are subtle differences in the sources of the two isotopes (mass 
and metallicity of
star, depth within star and effect of the final explosion) which make the relative
distributions of $^{60}$Fe and $^{26}$Al a potential source of information
on fine details of massive-star evolution.

In this paper we report the detection of two of the $^{60}$Fe lines by the SPI
instrument, part of the {\em INTEGRAL\/} mission.  The significance of
this detection is not sufficient for us to draw any conclusions about the sky
distribution relative to $^{26}$Al.  However the mission is expected to continue 
for several years, by which time there may be enough data for
spatial information to be extracted.  Earlier measurements (summarized by 
Harris et al. 1997)
did not detect $^{60}$Fe, and yielded only upper limits on the $^{60}$Fe/$^{26}$Al
ratio, until the preliminary detection reported 
by {\em RHESSI\/} (Smith 2004), which is consistent with ours (\S 3.4).

\section{Observations and analysis}

The {\em INTEGRAL\/} spacecraft was launched October 17 2002 into a high-inclination,
high-eccentricity orbit with a 3-day period.  It carries
two major $\gamma$-ray instruments, the co-aligned
IBIS and SPI.  Although each performs both imaging and spectroscopic tasks, IBIS is
designed for fine spatial resolution while SPI has superior energy resolution.
Its 19 Ge detectors achieve $\sim 0.3$\% resolution around 1 MeV; imaging at
the level $\sim 3^{\circ}$ within a $16^{\circ} \times 16^{\circ}$ field is 
enabled by a coded mask permitting differential illumination of the detectors as
a function of angle.  In our analysis we do not make use of this capability, 
because of the weakness of the lines.   

One input into our analysis is therefore an assumption about the Galactic
distribution of $^{60}$Fe.  We used the distribution of far-infrared (240 $\mu$m)
emission mapped by {\em COBE\/}/DIRBE (Hauser et al. 1997) which is expected to be a good
guide to the distribution of massive stars and is in fact one of the best predictors of
$^{26}$Al emission (Kn\"{o}dlseder et al. 1999).  The observations
available for use in our analysis were made during the first year of operations
(orbits 19--130) and largely came from two core (instrument team) programmes 
--- two deep exposures (4 Ms) to the Galactic centre, and a periodic scan
along the Galactic plane.  We also made use of data publicly released as of 2004
December.
This amounted to 13.5 Ms distributed over the whole sky, but favouring the inner Galaxy
locations where the DIRBE 240 $\mu$m and COMPTEL $^{26}$Al maps and the massive
star population peak (see e.g. the exposure map of Kn\"{o}dlseder et al. 2005).

During orbits 19--130 SPI's performance was nominal, with all 19 detectors
operating at full efficiency (effective area 70 $\hbox{cm}^2$) and resolution (2.5
keV FWHM at 1.3 MeV).  The energy resolution was maintained against the degradation
caused by cosmic-ray impacts by in-orbit annealing, which was effective in restoring
performance.  In our analysis we allowed for the fact that comparisons between spectra taken at
different epochs would have found slightly different widths and energies for the same line,
due to this time variability of resolution (\S 3.2).  Similarly, we analyzed separately
single-detector events (SE) and ``multiple"
events (ME) where the $\gamma$-ray energy was deposited in two detectors because
their effective spectral resolution and instrumental backgrounds differ somewhat.

\begin{figure}[tbh]
\vspace{-3mm}
\epsfxsize=\hsize % 8.6truecm
\epsfbox[54 365 558 710] {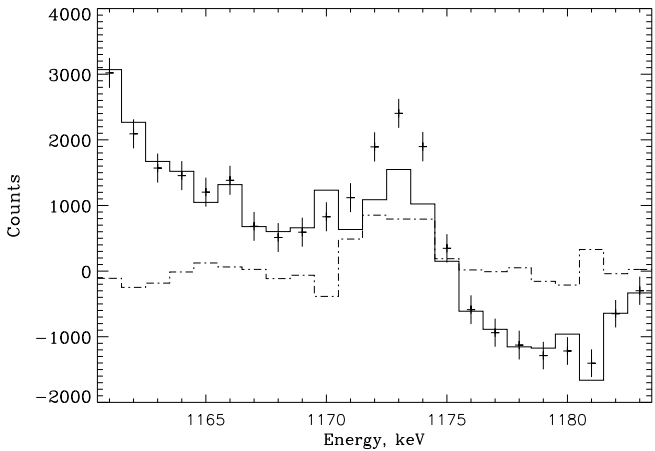}
\caption{Amplitudes of the GEDSAT background term (full line) and the 
GEDSAT $\bigotimes$ 7.6 yr background term (dot-dash line) fitted (along with
the $^{60}$Fe map exposure) to the time series of SE count rates
for 1161--1184 keV.  Data points --- background counts in 1 keV bins
with flat continuum subtracted.}
\end{figure}

The basic principle behind our analysis was the creation of models describing the
time variation of the background count rates, which dominate the cosmic signal by a
factor $\sim 100$.  The models were made up from physically
plausible environmental quantities that ought to contribute to the background.
Empirically, the prompt component of the background (both line and
continuum) is well described by the count rates in the Ge detectors when
saturated (energy loss $>8$ MeV, hereafter ``GEDSAT" rates).  But 
the prompt interactions also create radioactive isotopes with finite half lives,
which are strong sources of background lines.  The most important
example is radioactive $^{60}$Co, which decays with a 7.6-year
mean life emitting two $\gamma$-ray lines identical with those from cosmic $^{60}$Fe.
We modelled this time series by convolving the source term (GEDSAT) with
an exponential increasing on a 7.6 yr time scale.  We define the term {\em template\/}
to mean a function of environmental variables whose time series we will fit
to the background data, whether simple (like GEDSAT) or complex (such as 
GEDSAT $\bigotimes$ 7.6 yr).  If there is a cosmic source of the lines, it should follow 
a quite different time series (SPI's successive exposures to the 240 $\mu$m map),
so we make this the third term in the fit.

The best results were obtained when all three components (two background templates and
the expected $^{60}$Fe flux) were fitted simultaneously to the count rates 
by detector and pointing in 1 keV bins (\S 3.1).  We performed  
an alternative analysis in which the ``off-pointings" which are empty of $^{60}$Fe
according to the 240 $\mu$m map were
treated separately (\S 3.2).  The results of this are used only as a check,
and are not included in our final result, since they contain several systematic errors.
Finally, since the cosmic $^{60}$Fe/$^{26}$Al ratio is perhaps the most interesting
quantity which we can derive, we have analysed the 1.809 MeV line of $^{26}$Al in 
the identical way, so as to eliminate systematics from
the ratio if possible (\S 3.3).

\begin{figure}[tbh]
\vspace{-3mm}
\epsfxsize=\hsize % 17.2truecm
\epsfbox[54 365 558 710]{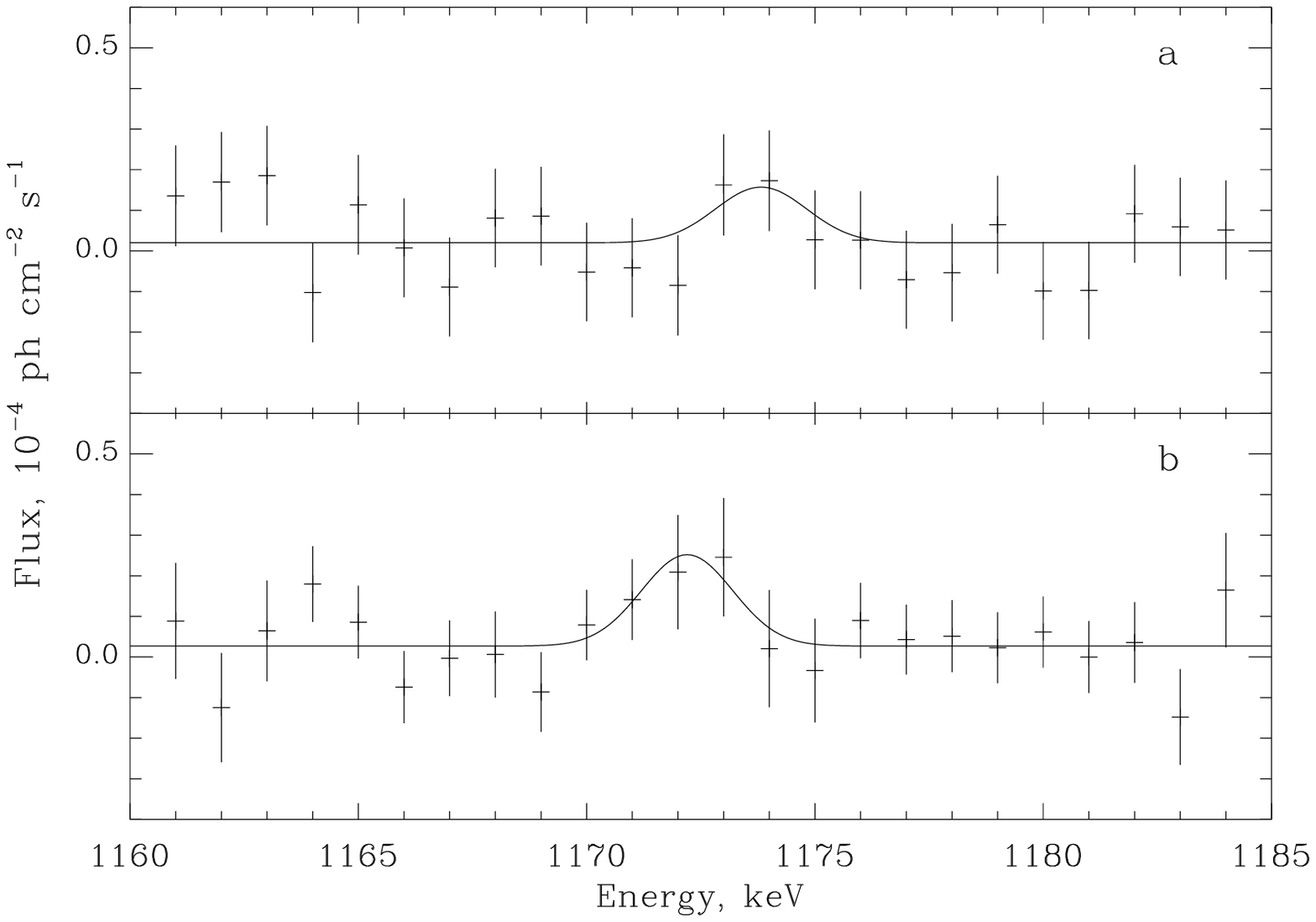}
\caption{Spectra of (a) SE and (b) ME from the analysis of
\S 3.1, with fitted $^{60}$Fe lines of strength 
$3.4 \pm 2.5 \times 10^{-5} ~\gamma ~ \hbox{cm}^{-2} \hbox{s}^{-1}$ and 
$5.6 \pm 2.7 \times 10^{-5} ~\gamma ~ \hbox{cm}^{-2} \hbox{s}^{-1}$ respectively.}
%\end{figure}

%\begin{figure}[tbh]
\epsfxsize=\hsize % 17.2truecm
\epsfbox[54 365 558 710]{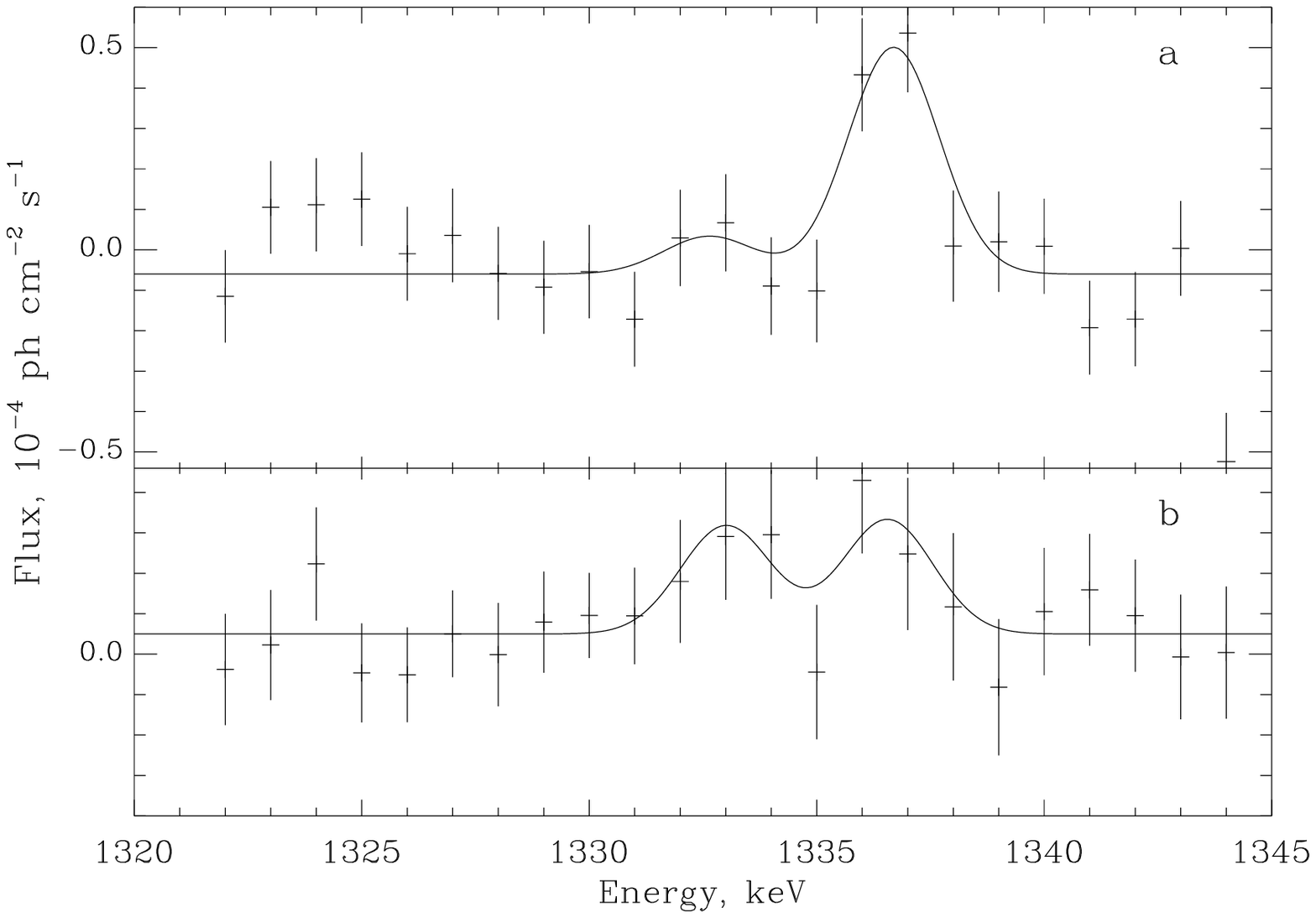}
\caption{Spectra of (a) SE and (b) ME obtained 
from the analysis of \S 3.1, with fitted $^{60}$Fe lines of strength
$2.3 \pm 1.6 \times 10^{-5} ~\gamma ~ \hbox{cm}^{-2} \hbox{s}^{-1}$ and 
$6.7 \pm 3.2 \times 10^{-5} ~\gamma ~ \hbox{cm}^{-2} \hbox{s}^{-1}$ 
respectively.  The blending $^{69}$Ge decay line is also fitted.}
\end{figure}
 
\section{Results and evaluation}

\subsection{The templates fitted to all pointings simultaneously}

In this analysis the count rates in the $^{60}$Fe lines were measured
as described below in each of the 
6821 pointings in our data set, and fitted by the templates GEDSAT and GEDSAT 
$\bigotimes$ 7.6 yr, and by the expected cosmic $^{60}$Fe flux (i.e. SPI's exposure
to the 240 $\mu$m map).  A template for the continuum which consisted of GEDSAT convolved
with the time series of count rates summed over two bands on either side of each line 
(1163--1169, 1177--1183 keV, and 1318--1328, 1349-1359 keV)
was kept fixed in the fit.  The line count rates to be fitted were obtained by
summing over the intervening intervals (avoiding blending lines as in 1335--1348 keV).
The line count rates in 1 keV bins, and the best-fitting amplitudes of GEDSAT and
GEDSAT $\bigotimes$ 7.6 yr, are shown for SE data in Fig. 1.  Clearly
the variation of the count
rates away from the line is explained by high amplitudes of the GEDSAT variable,
while those in the line itself are strongly influenced by the 7.6 yr
time-scale in the other template, which otherwise has amplitude zero as expected. 

The method thus explains the background (typically $\sim 99\%$ of the total) well, 
and we plot the amplitudes of the $^{60}$Fe term in the fit (signal-to-noise $\sim
1\%$) to get our result (Figs. 2--3).
The spectra appear to be quite free of systematics, except for the strong $^{69}$Ge
L-shell-capture decay line at 1337 keV.  A more subtle systematic error (\S 2)
might arise in the line energies and widths in Figs. 2 and 3 from the variation of the 
energy resolution (due to degradation and annealing) 
during the period covered by the measurements, which we 
consider below (\S 3.2c and footnote 1).  With   
this caveat we find the mean strength of each of the $^{60}$Fe lines to be 
$3.7 \pm 1.1 \times 10^{-5} ~\gamma ~ \hbox{cm}^{-2} \hbox{s}^{-1}$
by fitting the spectra in Figs. 2 and 3 by a model consisting of either one or two
Gaussian lines of fixed width 2.4 keV FWHM plus a flat continuum.\footnote{ The
value 2.4 keV was obtained from measurements made after a correction had been
applied for the variability of the instrumental energy resolution (\S 3.2c).}  The
significance is best visualized by summing all four lines together (Fig. 4).

\subsection{The templates fitted to off-pointings}

In this alternative analysis method we attempted to derive a ``universal" combination
of templates which describes the variation of the $^{60}$Fe count rates during
those pointings for which we are fairly confident that there is no signal, i.e.
``off"-pointings towards Galactic latitudes $>20^{\circ}$.  Only background templates
were fitted to these data.  To the optimum combined template we applied a
correction for the
discrepancies in energy resolution due to detector degradation effects, which must exist between
the off-pointings and the Galactic pointings, using the algorithm
described by Kn\"{o}dlseder et al. (2004).  

We could then obtain a measurement of the $^{60}$Fe lines by fitting the fixed combination of
templates and expected cosmic line strengths to the Galactic pointing data.  
An example of the results is shown in Fig. 5.
There are clearly systematic effects due to the failure of the template to remove 
the time series of blending lines with various half-lives, notably 
$^{62}$Ni (1173 keV, prompt), $^{52}$Mn (1334 keV, $\tau = 8.066$ d)
and $^{69}$Ge (1337 keV, $\tau = 2.348$ d).
Even the inclusion in the combined template of convolutions of these lifetimes with GEDSAT
did not remove these lines.  We made the following corrections for such
systematic errors: \\
(a) The 1173 keV $^{62}$Ni line by chance happens to follow immediately
in the $^{62}$Ni de-excitation cascade after a transition at 1163 keV which is also
visible in our spectrum.  The branching ratio is 100\%, and the line strengths
are in the ratio 1:1 if the $^{62}$Ni is produced by spallation.  Measuring the
1163 keV line strength immediately gives the 1173 keV line strength to be
subtracted from the $^{60}$Fe line (dashed line in Fig. 5).  \\
(b) The $^{52}$Mn line strength cannot be deduced as in (a), so we
fitted the 1332.5 keV $^{60}$Fe line under the assumptions that the $^{52}$Mn line
flux was either free or set to zero, the difference in the $^{60}$Fe line flux 
being the systematic error.  \\
(c) The effect of the correction for SPI's variable energy resolution was measured
by fitting the spectra with, and without applying it.  When we fixed the
widths at the typical corrected value 2.4 keV, the differences between the 
fluxes measured were very small 
($\le 0.3 \times 10^{-5} ~\gamma ~ \hbox{cm}^{-2} \hbox{s}^{-1}$).

The result of this version of the analysis is
$4.0 \pm 1.1 (stat.) \pm 0.7 (syst.) \times 10^{-5} ~\gamma ~ \hbox{cm}^{-2} \hbox{s}^{-1}$

\subsection{Measurement of the 1.809 keV line of $^{26}$Al}

A measurement of the 1.809 $^{26}$Al line by exactly the same method as in 
\S 3.1 yielded a result
$3.4 \pm 0.2 \times 10^{-4} ~\gamma ~ \hbox{cm}^{-2} \hbox{s}^{-1}$,
for a $^{60}$Fe/$^{26}$Al line flux ratio of $0.11 \pm 0.03$, corresponding
to an abundance ratio $0.23 \pm 0.08$.

\subsection{Evaluation}

The close agreement between the results of the two analyses (\S \S 3.1, 3.2)
suggests that the correction to the line widths and 
energies for the varying instrumental resolution
probably has little effect in \S 3.1.  In view of this lack of
systematic errors, we regard this as our best result, i.e.
$3.7 \pm 1.1 \times 10^{-5} ~\gamma ~ \hbox{cm}^{-2} \hbox{s}^{-1}$, with a 
possible systematic error $\pm 0.3 \times 10^{-5}$ due to the non-uniform
energy resolution.  

The significance $\sim 3 \sigma$ is rather better than that of Smith's 
(2004) result, $3.6 \pm 1.4 \times 10^{-5} ~\gamma ~ \hbox{cm}^{-2} \hbox{s}^{-1}$,
but the agreement between the two is very good.

\begin{figure}[tbh]
\vspace{-3mm}
\epsfxsize=\hsize % 17.2truecm
\epsfbox[54 365 558 710]{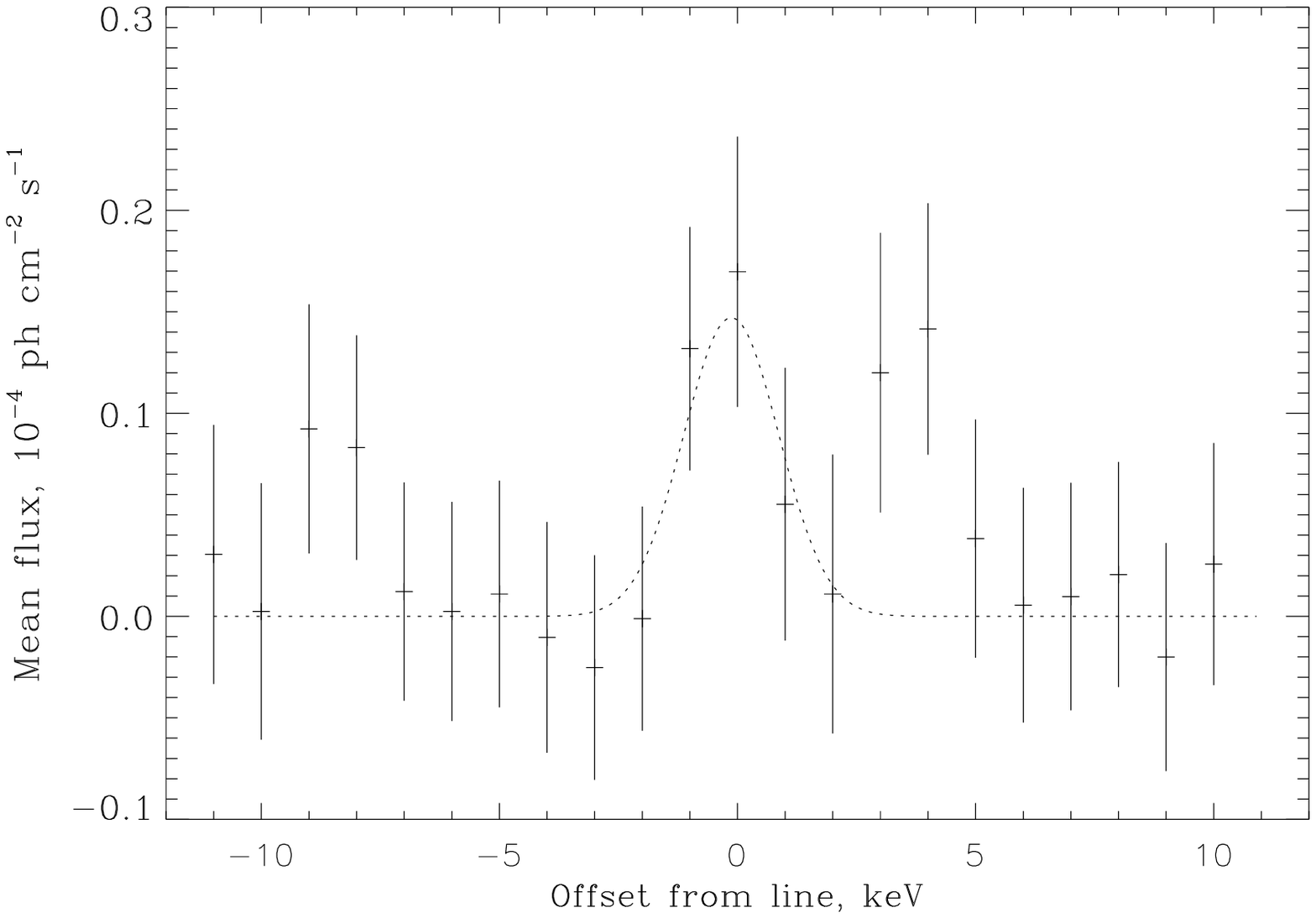}
\caption{Spectra of Figs. 2 and 3 overlaid and summed to give the total
$^{60}$Fe signal.  Dotted line --- the 
$3.7 \times 10^{-5} ~\gamma ~ \hbox{cm}^{-2} \hbox{s}^{-1}$ line deduced from
combining Figs. 2 (a,b) and 3 (a,b).}
%\end{figure}

%\begin{figure}[tbh]
\epsfxsize=\hsize % 8.6truecm
\epsfbox[54 365 558 710]{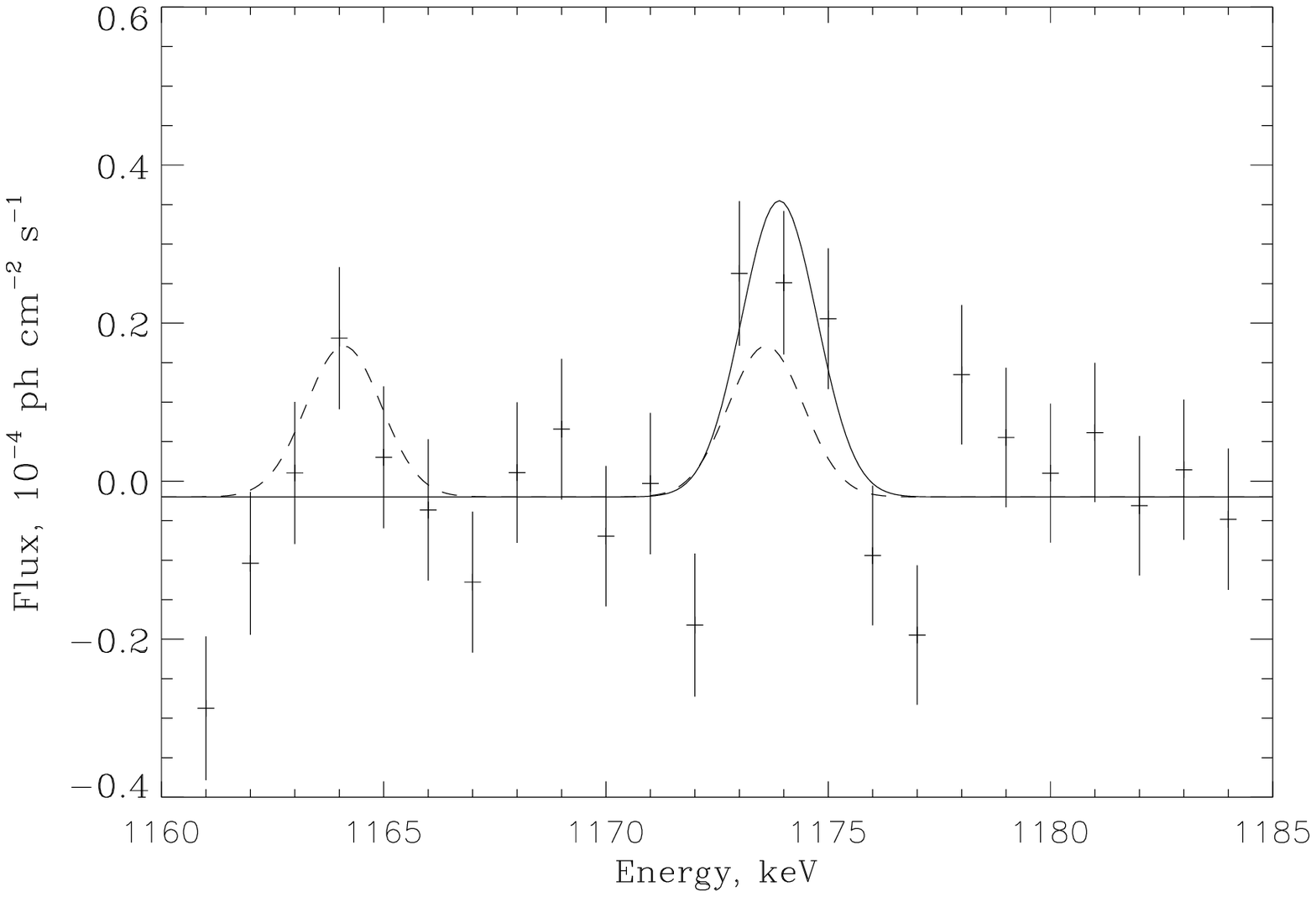}
\caption{Spectrum of SE obtained by the analysis of \S 3.2.
Full line --- fitted $^{60}$Fe line of strength
$7.9 \pm 2.1 \times 10^{-5} ~\gamma ~ \hbox{cm}^{-2} \hbox{s}^{-1}$.
Dashed line --- fitted $^{62}$Ni line of strength
$4.0 \pm 1.9 \times 10^{-5} ~\gamma ~ \hbox{cm}^{-2} \hbox{s}^{-1}$ (1163
keV), and inferred line of equal strength at 1173 keV.}
\end{figure}

\section{Discussion}

  In the context of massive star evolution, $^{26}$Al comes from zones 
containing free protons (H and Ne burning) while $^{60}$Fe requires a substantial free
neutron abundance (C, Ne and to a small extent He burning).  In the final supernova
explosion, they will be produced in roughly equal amounts (Limongi \& Chieffi 
2003).  Prantzos (2004) pointed out 
the contradiction between this expected CCSN ratio and results such as ours and
Smith's (2004), where the ratio is $\sim 0.2$.  His conjecture that there is a large
additional source of $^{26}$Al which acts prior to the core collapse and explosion
appears to be borne out by the models of Palacios et al. (2005), who find this 
source to be the massive winds expelled during the Wolf-Rayet (WR) phase.  The key point
is that there is a large abundance of $^{26}$Al in H-burning layers which are close enough 
to the surface for the wind to expel it during the star's short presupernova life.
The $^{60}$Fe abundance is much further inside.

Surprisingly, therefore, the Galactic distributions of the $^{26}$Al and $^{60}$Fe
lines may be quite different.  WR stars differ from the average SNII progenitor in
being (a) somewhat more massive on average and (b) highly dependent on metallicity.
The $^{26}$Al map exhibits ``hot spots" in areas like Cygnus which are too young
for even their most massive stars to have become SNII, but in which WR winds are
already active (Kn\"{o}dlseder et al. 2002); $^{60}$Fe emission should not be seen 
from these regions.  The metallicity gradient in the Galaxy is substantial
enough for excess $^{26}$Al emission to be seen from the inner Galaxy in
COMPTEL data (Palacios et al. 2005); $^{60}$Fe should be more evenly distributed.
When the possibility is factored in that some $^{60}$Fe is made in SNIa 
(Iwamoto et al. 1999), and some $^{26}$Al in AGB stars and novae (Diehl 2001), which
have a quite different history and distribution,
it appears that we must expect the unexpected when the data become sufficient for
a $^{60}$Fe map to be made. 
   
\acknowledgements{The SPI project was completed under the responsibility and
leadership of CNES.
We are grateful to ASI, CEA, CNES, DLR, ESA, INTA, NASA and OSC for support.}


\begin{thebibliography}{}
\bibitem[2001]{diehl2001} Diehl, R. 2001, in: The Universe in Gamma
Rays, ed.: Sch\"{o}nfelder, V., Springer, Berlin, 233 
\bibitem[1997]{harris1997} Harris, M. J., Purcell, W. R.,
McNaron-Brown, K., et al. 1997, in: Proc. of the Fourth Compton Symposium,
AIP Conf. Proc. 410, ed.: Dermer, C. D., Strickman, M. S., \& Kurfess, J. D.,
AIP, New York, 1079
\bibitem[1997]{hauser1997} Hauser M.G., Kelsall T., Leisawitz D., \&
Weiland J. (Eds.) 1997, COBE Ref Pub. No. 97-A (Greenbelt, MD: NASA/GSFC), available 
in electronic form from the NSSDC
\bibitem[1999]{iwamoto1999} Iwamoto, K., Brachwitz, F., Nomoto, K., et al. 1999, ApJS, 125, 439
\bibitem[1999]{knoedlseder1999} Kn\"{o}dlseder, J., Bennett, K., Bloemen, H., et al. 1999, A\&A, 344, 68
\bibitem[2002]{knoedlseder2002} Kn\"{o}dlseder, J., Cervi\~{n}o, M., Le Duigou, 
J. M., et al. 2002, A\&A, 390, 945 
\bibitem[2004]{knoedlseder2004} Kn\"{o}dlseder, J., Valsesia, M., Allain, M., et al. 
2004, in: The {\em INTEGRAL\/} Universe (ESA
Special Publication 552), ed.: Sch\"{o}nfelder, V., Lichti, G. G., \& Winkler, C. 
(Noordwijk: ESA), 33
\bibitem[2005]{knoedlseder2005} Kn\"{o}dlseder, J., Jean, P., Lonjou, V., et al. 2005, submitted to A\&A
\bibitem[2003]{Limongi2003} Limongi, M., \& Chieffi, A. 2003, ApJ, 592, 404
\bibitem[1982]{mahoney1982} Mahoney, W. A., Ling, J. C., Jacobson, A. S., \& Lingenfelter, 
R. E. 1982, ApJ, 262, 742
\bibitem[2005]{palacios2005} Palacios, A., Meynet, G., Vuissoz, C., et al. 2005, A\&A, 429, 613
\bibitem[2002]{prantzos2004} Prantzos, N. 2004, A\&A, 420, 1033
\bibitem[2004]{smith2004} Smith, D. M. 2004, in: The {\em INTEGRAL\/} 
Universe (ESA Special Publication 552), ed.: Sch\"{o}nfelder, V., Lichti, 
G. G., \& Winkler, C. (Noordwijk: ESA), 45 $\equiv$ LANL preprint 
{\em astro-ph}/0404594



\end{thebibliography}
\end{document}